\documentclass[aps,pra,twocolumn,showpacs,amsmath,amssymb,superscriptaddress]{revtex4-1}

\usepackage{graphicx}
\usepackage{color}

\newcommand{\reffig}[1]{Fig.~\ref{#1}}
\newcommand{\refeq}[1]{Eq.~(\ref{#1})}

\begin{document}

\title{Stability of solitary waves in random nonlocal nonlinear media}
\author{F. Maucher}
\affiliation{Max Planck Institute for the Physics of Complex Systems, 01187 Dresden,
Germany}
\affiliation{Laser Physics Centre, Research School of
Physics and Engineering, Australian National University, Canberra, ACT
0200, Australia}
\author{W. Krolikowski}
\affiliation{Laser Physics Centre, Research School of
Physics and Engineering, Australian National University, Canberra, ACT
0200, Australia}
\author{S. Skupin}
\affiliation{Max Planck Institute for the Physics of Complex Systems, 01187 Dresden,
Germany}
\affiliation{Friedrich Schiller University, Institute of Condensed Matter Theory and
Optics, 07743 Jena, Germany}

\begin{abstract}
We consider the interplay between nonlocal nonlinearity and randomness for two different nonlinear
Schr¨odinger models.We show by means of both numerical simulations and analytical estimates that the stability
of bright solitons in the presence of random perturbations increases dramatically with the nonlocality-induced
finite correlation length of the noise in the transverse plane. In fact, solitons are practically insensitive to noise
when the correlation length of the noise becomes comparable to the extent of the wave packet. We characterize
soliton stability using two different criteria based on the evolution of the Hamiltonian of the soliton and its
power. The first criterion allows us to estimate a time (or distance) over which the soliton preserves its form. The
second criterion gives the lifetime of the solitary wave packet in terms of its radiative power losses. We derive a
simplified mean field approach which allows us to calculate the power loss analytically in the physically relevant
case of weakly correlated noise, which in turn serves as a lower estimate of the lifetime for correlated noise in
the general case.
\end{abstract}
\maketitle

\section{Introduction}
In diverse physical settings the dynamics of nonlinear wave packets is
described by the nonlinear Schr\"odinger equation (NLS)~\cite{Sulem:book:1999},
including, for instance, nonlinear optics~\cite{Agrawal:book:2006},
Bose-Einstein condensates (BEC)~\cite{BEC_review, BEC_review2,Dalfovo:1999}, and water waves~\cite{Zakharov:1968}. More recently, the NLS equation has been also considered in the context of so-called
rogue waves~\cite{Akhmediev:PLA:2009}. The interplay of
diffraction or dispersion, respectively, which which naturally tends
to spread the wave,  and nonlinearity as governed by the NLS can lead to the formation of solitons solitons, i.e., robust localized
particle-like wave packets that do not change their shape upon
temporal evolution or spatial propagation~\cite{Chiao:64:PRL}.

Solitons are
ubiquitous in nature and can be found in many nonlinear systems ranging from
optics~\cite{Agrawal:book:2006}, physics of cold matter~\cite{Strecker:02:nature,Strecker:NJP:2003} and plasma~\cite{Porkolab:1976,Davydova:1998} to
biology~\cite{Davydov_soliton}.
In realistic settings nonlinear systems supporting solitons are often
subject to random perturbations~\cite{Konotop_book}. Such perturbations may  arise from the
fluctuation of the external linear potential confining  the wave, as in the case of BEC's in
spatially and temporally fluctuating trapping potentials~\cite{Cockburn:PRL:2010,Lye:PRL:2005}, optical beams in
nonlinear dielectric waveguides~\cite{Gaididei:OL:1998}, or waveguide arrays~\cite{Pertsch:PRL:2004} with
random variation of refractive index,   size, or  waveguide spacing. Furthermore, the optical nonlinearity of
nematic liquid crystals can exhibit stochastic variation due to
fluctuations of the crystal temperature or conditions of surface anchoring (e.g., roughness) affecting the orientation of the
crystal's molecules~\cite{Conti:PRE:2005,Barberi:PRA:1990}.
Similarly, fluctuations of temperature will introduce randomness in
colloidal suspensions~\cite{Reece:PRL:2007,Conti:PRL:2006}, while noise in the magnetic field
employed to control scattering length of the BEC via Feshbach resonance~\cite{Chin:RevModPhys:2010}
will  result in the stochasticity of its nonlinear interaction potential~\cite{Lewenstein:Nature:2010}.
Also, stochasticity in  the Gross-Pitaevskii equation must be often taken into account to describe quantum effects in dilute ultra-cold Bose-gases (e.g.,~\cite{Blakie:AiP:2008}).
Finally, fabrication-induced spatial fluctuations  in periodic ferroelectric domain patterns in quadratic
media act as  a source of spatial disorder  in the nonlinearity of the quasi-phase matched parametric wave
interaction affecting the propagation of quadratic solitons~\cite{Torner:JOSAP:97,Clausen:OL:97,Conti:OL:2010}.

It has been well appreciated  that randomness in the linear or nonlinear  potential supporting solitons
may have dramatic consequences on their stability and dynamics
depending on the strength of disorder. The presence of randomness leads to radiation being emitted by the
self-guided wave packet, the amount of which depends crucially on the typical length
scale of the fluctuation or correlation of the noise. The emission of radiation weakens the
self-induced localization and, ultimately leads to the decay of solitons. In fact, it has been shown
that disorder is equivalent to the presence of  an effective loss in the nonlinear system~\cite{Clausen:OL:97,Abdullaev_book, Conti:OL:2010}.
On the other hand, it appears that  the interplay between nonlinearity and weak randomness can
lead to diverse interesting phenomena, such as random walk of solitons in the transverse
plane~\cite{Gordon:OL:86,Kartashov:PRA:2008,Conti:PRL:2011} or
Anderson localization~\cite{Pertsch:PRL:2004,Inguscio:Nature:2004,
Lahini:PRL:2008,Inguscio:Nature:2008,Lewenstein:Nature:2010}.
Up to now, mostly local nonlinear
interaction has been considered in studies of solitons in nonlinear random systems.
This amounts to Kerr-type nonlinear optical response and contact boson interaction in BEC.
Recently however, a few works appeared dealing with random systems that exhibit spatially nonlocal
nonlinearity~\cite{Conti:PRE:2005,Conti:PRL:2006,Conti:PRL:2011,Peschel:PRA:2011}.

Nonlocality of the  nonlinear response appears to be common to a great variety of nonlinear systems.
Physically speaking, nonlocality means that the nonlinear response of the medium
in a specific location  is determined by the wave amplitude in a certain neighborhood of this location. The extent of this neighborhood is
often referred to as the degree of nonlocality. Nonlocality
is common to media where certain transport processes such as heat or charge transfer~\cite{Litvak:1975},
diffusion~\cite{Suter:pra:1993} and/or drift~\cite{Saffman:prl:2007} of atoms
are responsible for the nonlinearity. It also occurs in media with long-range inter-particle interaction.
This is the case of nematic liquid crystals where nonlinearity involves the 
reorientation of induced dipoles~\cite{Conti:prl:2004}   and in the context
of BECs with non-contact long-range
interatomic interaction \cite{Goral:PRA:05,Lahaye:RepProgr:09,Maucher:PRL:2011}.
Nonlocality of nonlinearity and its impact
on solitons has been studied extensively in the last decade. One of the most important features of
nonlocality is its ability  to  arrest catastrophic collapse of multidimensional waves~\cite{Turitsyn:tmf:85,Bang:pre:2002,Lushnikov:PRA:2010,Maucher:nonlinearity:2011}
and stabilize complex solitonic structures~\cite{Briedis:opex:2005,Lashkin:pra:2007,Skupin:pre:2006,Maucher:2010:PRA}.
These stabilizing properties of nonlocality have also been identified in the presence of randomness.
For instance, in recent studies of  many-soliton interaction in disordered nonlocal medium Conti {\em et al.} have
demonstrated that nonlocality leads to the formation of soliton clusters
and noise quenching~\cite{Conti:PRE:2005,Conti:PRL:2006}. Batz {\em et al.}~\cite{Peschel:PRA:2011}
reported  nonlocality-mediated  decrease of the  quantum phase diffusion and increased
coherence of quantum solitons while  Folli {\em et al.}~\cite{Conti:PRL:2011} have shown that the soliton
random walk can be efficiently suppressed in highly nonlocal media.

In the present work we will study the effect of nonlocality on the stability of solitons in nonlinear random media.
While many previous papers consider only ``longitudinal'' random perturbations [i.e., a situation
where the randomness is
only a function of the longitudinal (propagation) coordinate (e.g.~\cite{Abdullaev_book})], we will
deal here with the general case when randomness is a function of both propagation and transverse coordinates.
We will consider two practically relevant models for  random nonlocal systems. In the first model
the randomness contributes additively towards the nonlinear response of the medium.
In the second model, randomness directly affects the parameters characterizing the nonlinear response of the medium (i.e., the noise itself becomes nonlinear).
We will show that nonlocality  stabilizes solitons by effectively increasing the correlation length of the random perturbation. By using a simplified mean field approach we are able to give a lower bound for the lifetime of solitary wave packets in the case of weakly correlated noise.

The paper is organized as follows:
In Sec.~\ref{sec:model}, we will introduce the afore mentioned nonlocal model with additive noise.
This model naturally incorporates the interplay between randomness and nonlocality,
and the noise term in the field equation is linear.
The different effects of randomness and nonlocality on the soliton dynamics will be studied first separately in Secs~\ref{sec:noise} and \ref{sec:nonlocal}, and for the full model in Secs.~\ref{sec:full}.
In Sec.~\ref{sec:nlnoise} we investigate our second random nonlocal model, where the noise acts multiplicatively and randomness becomes nonlinear as well. Finally, we will conclude in Sec.~\ref{sec:conclusion}.

\section{Nonlocal model with additive noise}\label{sec:model}
We will consider the evolution of the wave function  $\psi(x,t)$  with $x$ and $t$
denoting generalized transverse and longitudinal (propagation) coordinates,
respectively. The function  $\psi$ may represent the main electric field component  of a linearly polarized light beam or the wave function of a quantum object such as BEC. We assume that $\psi$ satisfies the following system of coupled equations
\cite{Conti:PRE:2005,Conti:PRL:2011}:
\begin{subequations}
\label{eq:model1}
\begin{align}
i\partial_t \psi(x,t) +\partial_{xx}\psi (x,t)+  \rho \psi(x,t) & = 0 \\
-\sigma^2\partial_{xx}\rho+\rho & = |\psi|^2 + \epsilon. \label{eq:model1b}
\end{align}
\end{subequations}
Here, $\rho$ represents the nonlinear response of the medium. In the context of nonlinear optics, $\rho$ is usually
identified with a nonlinear refractive index change, while it would account for the effective two-body interaction potential in the case of a BEC.
In particular, \refeq{eq:model1} describes optical beam propagation in a planarly aligned nematic liquid crystal cell, which is known to exhibit a substantial nonlocal nonlinearity of molecular origin~\cite{Assanto:2003:jqe,Conti:PRL:2003}. In such a setup, the nematic director of the aligned liquid crystal gets an additional tilt by the field of the light beam, which directly translates in a change of the effective refractive index. In \refeq{eq:model1b}, we account for random fluctuations of the nematic director, i.e., the molecular orientation.
In fact, such random perturbations of the molecular orientation across the volume of the crystal can be introduced
by fluctuations of the temperature, or the anchoring of the molecules of the crystals at the boundaries~\cite{Barberi:PRA:1990}.
The stochastic term $\epsilon$ is assumed to be a $\delta$ correlated Langevin noise in
both longitudinal and transverse coordinates.
Then, the noise fulfills  $ \langle\epsilon\rangle=0$
and $\langle\epsilon(x,t)\epsilon(x^\prime,t^\prime)\rangle = n^2\delta(t-t^\prime)\delta(x-x^\prime)$, where
$$ \langle f \rangle = \lim\limits_{N\rightarrow\infty} \frac{1}{N} \sum\limits_{j=1}^N f_j$$ denotes ensemble averaging over different stochastic realizations $f_j$, and $n$ is the so-called coupling strength.
By solving \refeq{eq:model1b} via Fourier transform  the system can be written as a single nonlocal NLS equation
\begin{equation}
\begin{split}
 i\partial_t \psi(x,t) +\partial_{xx}\psi(x,t) &\\
 + \psi(x,t)\int\limits_{-\infty}^{\infty}R(x-x')\left[|\psi|^2(x',t) + \epsilon(x',t)\right]\mathrm{d}x' & = 0,
 \end{split}\label{eq:model1nls}
 \end{equation}
with the nonlocal response function
\begin{equation}
R(x)=\frac{1}{2\sigma}\mathrm{e}^{-\frac{|x|}{\sigma}}.
\label{eq:exponential}
\end{equation}
Because the noise term $\epsilon(x,t)$ is additive in the constituent equation \refeq{eq:model1b},
it acts as a source term affecting the medium independently of whether the actual signal
(e.g., the optical beam) is present or not.  Therefore, in the nonlocal NLS \refeq{eq:model1nls} noise plays the role of
a random background potential and is not affected  by the nonlinearity itself.
The parameter $\sigma$ represents the extent of the nonlocality of the  nonlinear response and
hence defines  different nonlocal regimes. Without the
noise term, \refeq{eq:model1} supports stable nonlocal solitons~\cite{Skupin:pre:2006}.

In \refeq{eq:model1}, the nonlocality parameter  $\sigma$ leads to both nonlocal nonlinearity
and finite correlation length.
To clarify the effect of each of these two constituents on the  stability of solitons we will first discuss both of them separately, and then consider their combined action.

\subsection{Effect of transverse correlation of the noise}\label{sec:noise}

In this subsection we will discuss the effect of transverse correlation of the randomness on  dynamics of {\em local} solitons.
To this end, we will consider the following NLS
\begin{equation}
 i\partial_t \psi +\partial_{xx}\psi+ | \psi |^2  \psi + \eta(x,t)  \psi = 0,
\label{eq:local_nls}
\end{equation}
where  the  random perturbation term is given by the nonlocal relation
\begin{equation}
\eta(x,t) = \int R(x-x^\prime) \epsilon(x^\prime,t)\mathrm{d}x^\prime.
\label{eta}
\end{equation}
Here, $\epsilon(x)$ is again a white noise.
For the sake of consistency with the original model \refeq{eq:model1}, we will use the exponential function \refeq{eq:exponential} as kernel function. However, we verified that our
findings also hold for other, e.g.~Gaussian, correlation functions.
Apart from being a straight forward simplification of \refeq{eq:model1}, \refeq{eq:local_nls} has numerous direct physical motivations. For example, the pioneering work by Gordon and Haus~\cite{Gordon:OL:86} on the random walk of optical solitons in fiber transmission lines involves the above local NLS.

\begin{figure}
\includegraphics[width=0.99\columnwidth]{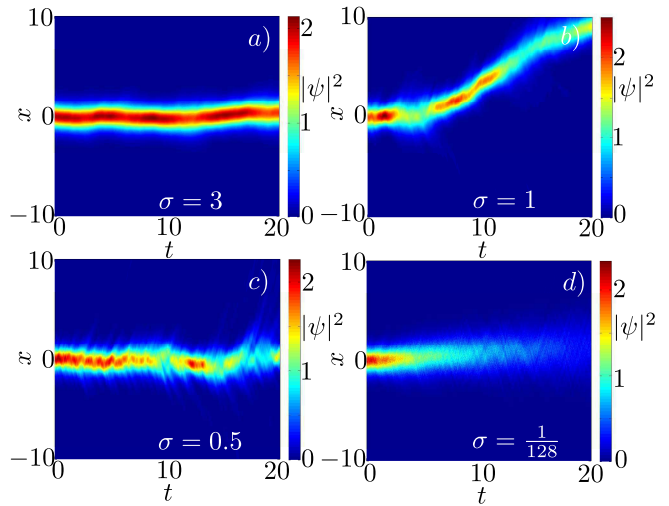}
\caption{(Color online) Single realizations of the propagation of local solitons perturbed by correlated
noise for different values of $\sigma$ (degree of nonlocality). In (a)-(d), we have
 $\sigma=3,$ 1, 0.5, $1/128$, with the respective coupling strength $n=\sqrt{\sigma/5}$.
The decay of the soliton due to the emission of radiation becomes stronger for smaller $\sigma$, even though the coupling strength is reduced as well.
Apart from the decay, solitons perform a random walk in the transverse plane.
}\label{fig1}
\end{figure}

The dynamics of localized solutions of \refeq{eq:local_nls} is determined by the ratio of two length scales, the width $\sigma$ of the response function $R$ in~\refeq{eta}
and the $1/e$-width
$\sigma_\mathrm{S}=\mathrm{arcosh}(\sqrt{e})/\sqrt{\lambda}\approx 1.085/\sqrt{\lambda}$ of the modulus squared of the initial soliton solution
\begin{equation}
 \psi_\mathrm{S}(x)=\frac{ \sqrt{2\lambda} } { \cosh{\sqrt{\lambda} x} }.
\label{eq:soliton_solution}
\end{equation}
Here, $\lambda>0$ is the so-called soliton parameter (or propagation constant) which determines the trivial phase evolution of the soliton in the unperturbed local NLS equation.
In the following, we will often use the width $\sigma_\mathrm{S}$ instead of $\lambda$ to characterize the soliton solutions, because in the nonlocal case we do not have an analytical expression like \refeq{eq:soliton_solution} for the soliton profile and have to resort to numerical solutions.
Figure~\ref{fig1} illustrates typical propagation scenarios of such solitons, here with initial condition $\lambda=1$ in~\refeq{eq:soliton_solution}, for succeedingly decreasing correlation length $\sigma$ and coupling-strength $n=\sqrt{\sigma/5}$
for single stochastic realizations.
Figure~\ref{fig1}(a) represents a realization for $\sigma\gg\sigma_\mathrm{S}$. One can see that in this case the peak intensity is only slightly modulated, and
insignificant radiation is produced.
A further decrease of   $\sigma$ [\reffig{fig1}(b)-(d)] leads to an evident  decay of the soliton due to the emission of radiation.
At the same time, the soliton itself performs a random walk in the transverse direction (Gordon-Hauss effect~\cite{Gordon:OL:86}).
The suppression of the soliton random walk by nonlocality has been already discussed by Conti {\em et al.}~\cite{Conti:PRL:2011}.
Here, we will focus on the noise-induced  decay of the solitons.

Since the original noise term $\epsilon$ is $\delta$ correlated, one can immediately find that
\begin{equation}
\langle\eta(x,t)\eta(x^\prime,t^\prime)\rangle=C(x-x^\prime)\delta(t-t^\prime). \label{eq:corr}
\end{equation}
For  $R$ given by \refeq{eq:exponential}, the function $C(x)$ reads
\begin{equation}
C(x) =\frac{n^2}{4\sigma^2} e^{-|x|/\sigma}(\sigma+|x|), \label{eq:autocorrelation}
\end{equation}
and describes the effect of nonlocality on the noise.
Namely, while the actual random source is represented by a white noise the nonlocal character of
the nonlinearity transforms it into  an {\em effective colored noise}. The nonlocality acts as a low-pass filter
eliminating the high frequencies of the original noise source and thereby smoothing out the randomness.
This is best appreciated in the spatial Fourier domain ($\tilde{f}=\mathcal{F}[f]$) where  where the white-noise perturbation is modified by a bandpass
filter defined by the Fourier spectrum of the nonlocal response function
\begin{equation}
\tilde{\eta}(k,t)=\sqrt{2\pi}\tilde{R}(k)\tilde{\epsilon}(k,t).
\end{equation}
As a result, the  effective noise $\eta$  exhibits a finite correlation
length determined by the degree of nonlocality $\sigma$.
The crucial dependence of the soliton dynamics on $\sigma$ is  shown in \reffig{fig1} for single realizations.
In the following we will investigate the behavior of averaged quantities $\langle\cdot\rangle$.

\subsubsection{Random phase shift}

Let us first consider the situation when the width
of the nonlocal response function $\sigma$ significantly exceeds  the spatial extent of the soliton (i.e., $\sigma \gg \sigma_\mathrm{S}$).
In this case, the  resulting  correlation length of the noise is large, so that the
dependence of the randomness
in the transverse plane is basically averaged out, and the noise $\eta$ only depends on the longitudinal
coordinate,  $\eta(x,t)=\eta(t)$.  Then, the noise term can be effectively removed from the original
stochastic \refeq{eq:local_nls} by the transformation $\hat{\psi}(x,t)=\psi(x,t)\exp[-i\int_{-\infty}^t\eta(t')dt']$.
In this regime of longitudinal-only disorder the soliton maintains its intensity profile $|\psi|^2$ while
acquiring a random phase shift. In other words, soliton amplitudes will evolve independently of the
strength of the noise.
However, the random phase shift of the soliton may become apparent in ensemble-averaged quantities, i.e., $\langle\psi\rangle$ decays in time, since each
realization acquires a different random phase shift upon propagation.

Generally, when the noise $\eta(x,t)$ depends on both transverse and longitudinal coordinates $x$ and $t$,
it is possible to show that (see App.~\ref{sec:FN} for details)
\begin{equation}
  \left[i\partial_t+\partial_{xx} + i \frac{C(0)}{2}\right]\langle \psi\rangle+  \langle| \psi |^2  \psi\rangle =  0.
\label{eq:local_mf}
\end{equation}
Unfortunately, due to the nonlinear term $\langle| \psi |^2  \psi\rangle$ it is not possible to solve \refeq{eq:local_mf} directly. However, for large correlation length $\sigma$
the perturbation to the soliton
due to spatial noise is weak and one can approximate $\partial_{xx} \langle \psi\rangle+  \langle| \psi |^2  \psi\rangle \approx \lambda \langle \psi\rangle$, where
 $\lambda$ is the soliton parameter from \refeq{eq:soliton_solution}. Then, \refeq{eq:local_mf} gives
\begin{equation}
 |\langle\psi\rangle|^2 \approx |\psi_\mathrm{S}|^2 e^{-C(0)t},
\label{eq:phase_decoherence}
\end{equation}
for times $t$ sufficiently small. Physically speaking, in \refeq{eq:phase_decoherence} we neglect random walk
and the  radiative decay of the soliton.

To verify the  above findings we must resort to numerical analysis.
The numerical methods we used throughout this paper are detailed in App.~\ref{sec:numerics}.
Due to the noise-induced emission of radiation
the absorbing boundaries we use emulate parts of the wave packet that leave the finite numerical
box and thus lead to an effective decrease of the total power (or number of particles) $P=\int|\psi|^2\mathrm{d}x$ in our simulations.
$P$ would remain a conserved quantity if we could use an infinitely large numerical box.
Generally, we assume that we can split the wave function into a solitonic part and a radiative part,
\begin{equation}
\psi = \psi_\mathrm{S}+\psi_\mathrm{R}.
\end{equation}
Then, the total power $P$ of the wave function can be decomposed into the corresponding partial powers
\begin{equation}
P=P_\mathrm{S}+P_\mathrm{R},
\end{equation}
and we have $P=P_\mathrm{S}=4\sqrt{\lambda}\approx4.34/\sigma_\mathrm{S}$ at $t=0$.

\begin{figure}
\includegraphics[width=0.99\columnwidth]{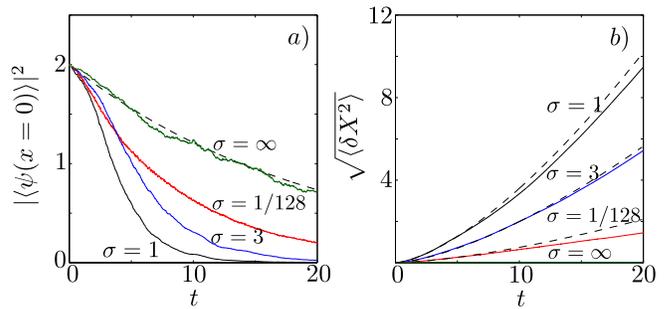}
\caption{(Color online) (a)  Evolution of $|\langle\psi(x=0,t)\rangle|^2$
(solid lines) compared to the analytical prediction
\refeq{eq:phase_decoherence} (dashed line).
The correlation length is $\sigma=\infty$, 3, 1, and 1/128 as indicated in the plot for the numerical solid lines. The auto-correlation $C(0)=n^2/4\sigma=1/20 $ is kept constant.
(b) The corresponding soliton random walk (solid lines) compared to the analytical prediction of \refeq{eq:randomwalk}  (dashed lines).
In our simulations, ensemble averages are taken over $128$ realizations.}
\label{fig2}
\end{figure}

Throughout this section we will consider $\psi(x)=\psi_\mathrm{S}(x)=\sqrt{2}/\cosh(x)$ ($\lambda=1$) as an initial condition at $t=0$.
Figure~\ref{fig2}(a) depicts the evolution of $|\langle\psi(x=0,t)\rangle|^2$ (solid lines), for different values of the correlation length $\sigma$.
The auto-correlation $C(0)=n^2/4\sigma$ is fixed to 1/20 by adjusting the coupling strength $n$, so that there is only one prediction (dashed line) for the evolution $|\langle\psi(x=0)\rangle|^2$ from \refeq{eq:phase_decoherence}.
As expected, we find that \refeq{eq:phase_decoherence} is exact
for infinite correlation length $\sigma=\infty$ (i.e., no spatial noise).
For finite values of $\sigma$, \refeq{eq:phase_decoherence} holds for small times $t$ only,
because of the combined effect of the emission of radiation and random walk. In fact, for large $\sigma>1$
it is mainly the random walk which causes the deviation of $|\langle\psi(x=0)\rangle|^2$
from the predictions of \refeq{eq:phase_decoherence}, whereas for small $\sigma<1$ radiative losses
to the soliton are more important (see also \reffig{fig1}).
The combined effect of these two effects leads to the breaking of the order of the curves in~\reffig{fig2}(a): the curve for $\sigma=1/128$, where radiation losses are strongest but random walk is small [see also~\reffig{fig2}(b) and the next section] - intersects the curves for $\sigma=1$ and $\sigma=3$.

\subsubsection{Random walk}

The
random walk of the solitons can be estimated analytically.
With $X(t)=\int x|\psi|^2\mathrm{d}x/P$, $\delta X(t)=X-\langle X \rangle$,
 it is possible to show that \cite{Conti:PRL:2011}
\begin{subequations}
\label{eq:randomwalk}
\begin{align}
\langle \delta X(t)^2 \rangle & = \frac{16 n^2 \kappa}{3P_\mathrm{S}^2}t^3 \\
\kappa & = \iiint \psi_\mathrm{S}(x_1) \left[\partial_{x_1}\psi_\mathrm{S}(x_1)\right]R(x_1-x_3) \label{eq:randomkappa}\\ & \quad \times \psi_\mathrm{S}(x_2) \left[\partial_{x_2}\psi_\mathrm{S}(x_2)\right] R(x_2-x_3)  \mathrm{d}x_1 \mathrm{d}x_2 \mathrm{d}x_3. \nonumber
\end{align}
\end{subequations}
In Fig.~\ref{fig2}(b) we compare predictions of \refeq{eq:randomwalk} (dashed lines) to numerical results (solid lines).
Whereas we find perfect agreement for large degree of
nonlocality $\sigma$, there is a clear deviation for smaller $\sigma$.
This is due to the increase of radiative losses for smaller $\sigma$'s. When
the soliton power decreases upon propagation, the random walk decreases as well.
To illustrate this behavior, let us consider the situation when $\sigma$ is significantly smaller
than the width  $\sigma_\mathrm{S}$ of the spatial
soliton under consideration. We see that according to \refeq{eq:randomwalk}
the strength of the random walk only depends on $n^2$, because $\kappa/P_\mathrm{S}^2\rightarrow\lambda^{3/2}/15$ for $\sigma \rightarrow 0$.
However, we have to take into account that in this regime radiation
losses to the soliton affect $\kappa/P_\mathrm{S}^2$ during propagation and introduce deviations from
$\langle \delta X(t)^2 \rangle \propto t^3$~\footnote{If we assume adiabatic transformation
into  solitons with lower $\lambda$ we find that $\kappa/P_\mathrm{S}^2 \propto \lambda^{3/2}$.}.

\subsubsection{Hamiltonian}

\begin{figure}
\includegraphics[width=0.99\columnwidth]{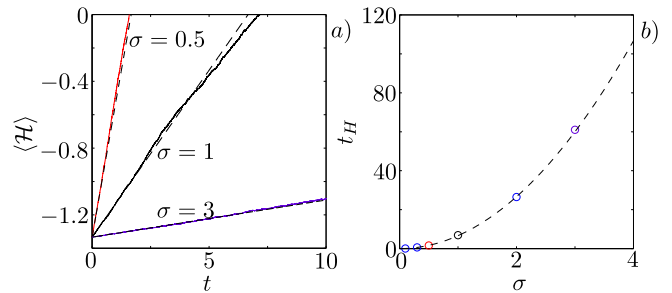}
\caption{(Color online) 
(a)  Evolution of Hamiltonian $\mathcal{H}$ [\refeq{hamlocal}]
(solid lines) compared to analytical predictions from
\refeq{eq:hamiltonian} (dashed lines).
The correlation length is $\sigma=3$, 1, and 0.5 as indicated in the plot. The auto-correlation $C(0)=n^2/4\sigma=1/20 $ is kept constant.
(b) The analytical prediction for $t_\mathcal{H}$ (dashed black line) as a function of $\sigma$ is shown.
Numerical values are indicated by circles, where the colored circles correspond to the colors used in (a). In our simulations, ensemble averages are taken over $128$ realizations.
}
\label{fig3}
\end{figure}

The evolution of the Hamiltonian can be seen as a measure of the deformation of the initial soliton profile due to random perturbations. Hence, it can used to characterize typical times upon which deviations from the input shape develop during propagation.
We introduce the Hamiltonian as
\begin{equation}
 \mathcal{H} = \int |\partial_x \psi|^2 -\frac{1}{2}|\psi|^4 \mathrm{d}x, \label{hamlocal}
\end{equation}
i.e., the Hamiltonian of Eq.~(4) when $\eta \equiv 0$.
Thus, in the limit $n\rightarrow 0$ the Hamiltonian $\mathcal{H}$ is a conserved quantity, but it
becomes time dependent for finite coupling strength.
Several papers~\cite{Fannjiang:JPhysA:2006,Fannjiang:physD:2005,Debussche:physD:2002} already emphasized the fact that $\langle\mathcal{H}\rangle $ is a linear function of time $t$,
$\langle\mathcal{H}(t)\rangle = \mathcal{H}_0 + \gamma t$.
In fact, it is possible to compute the ascent $\gamma$ analytically (see App.~\ref{sec:mean_ham}).
Then, the final result for the time evolution of the averaged Hamiltonian reads
\begin{equation}
\label{eq:hamiltonian}
 \langle \mathcal{H}\rangle = \mathcal{H}_0 - \frac{d^2 C(x)}{d x^2}\,\bigg|_{x=0} P t=\mathcal{H}_0 + \frac{n^2 P}{4\sigma^3} t.
\end{equation}
This formula emphasizes the huge impact of the nonlocality mediated correlation length $\sigma$ on the soliton propagation:
the ascent $\gamma$ scales with $1/\sigma^3$, whereas the noise amplitude enters only as $n^2$ [see also \reffig{fig3}(b)].
In order to define a time upon which the initial soliton profile does not change its shape significantly, we have to choose a ''critical value'' of the mean Hamiltonian.
This choice is rather arbitrary, depending on how large deviations from the input shape are considered to be significant.
Here, we assume that the soliton profile remains almost unchanged until the time $t_\mathcal{H}=-\mathcal{H}_0/\gamma$ is reached, i.e.,
$\langle\mathcal{H}(t_\mathcal{H})\rangle=0$ and $\gamma=n^2 P/4\sigma^3$. As we observe in numerical simulations, after time $t_\mathcal{H}$  only a small fraction of power of the soliton is converted into
radiation, and the soliton is still a perfectly well localized wave packet.
In Fig.~\ref{fig3}(a) we compare predictions of \refeq{eq:hamiltonian} (dashed lines) to numerical results (solid lines). As long as the total power $P$ is  conserved quantity in the simulations, i.e., the numerical box is large enough, we find perfect agreement. However, noise generated radiation spreads quite fast and will eventually reach any numerical boundaries.
Even though we used numerical boxes of size 300-600 and 8000-16000 points in $x$, for larger propagation times $t$ numerical curves deviate from the ideal linear behavior.

\subsubsection{radiative losses}

As we have seen so far, the effect of emission of radiation and subsequent loss of power is crucial for the
soliton dynamics, particularly when $\sigma$ is small compared to $\sigma_\mathrm{S}$. 
In the following, we will refer to soliton stability in terms of the  decay or decrease of the peak intensity
$|\psi|^2_{\rm max}(t)=\max_{x}|\psi|^2 $ upon propagation.
Due to the stochastic character of the propagation \refeq{eq:local_nls} we will
discuss, in fact, the ensemble averaged quantity $\langle|\psi|^2_{\rm max}\rangle$.

It is actually possible to estimate analytically the radiative losses to the soliton in the limit $\sigma\ll\sigma_\mathrm{S}$.
To this end, let us assume that the noise term $\eta\psi$ in \refeq{eq:local_nls} acts perturbatively (of the order $\delta\ll1$)
on the soliton $\psi_\mathrm{S}$, and $\psi=(\psi_\mathrm{S}+\chi)\exp(i \lambda t)$, where $\chi(x,t)$ is of the order $\delta\ll1$ as well. Then, in order $\delta^1$ we find
$$ i\partial_t \chi - \lambda \chi + \partial_{xx}\chi + 2 |\psi_\mathrm{S}|^2 \chi + \psi_\mathrm{S} ^2\chi^* +\eta\psi_\mathrm{S} = 0. $$
At initial time $t=0$ we have the pure soliton $\psi_\mathrm{S}$ without any perturbation, and thus $\chi(x,t=0)=0$.
For small times $\Delta t$ we can therefore write down a formal solution for the perturbation
\begin{equation}
\chi(x,\Delta t) \approx i\int\limits_0^{\Delta t} \eta(x,t)\psi_\mathrm{S}(x)\mathrm{d}t.
\label{eq:formalpert}
\end{equation}
When the correlation length $\sigma$ of the noise is much smaller than the width of the soliton $\sigma_\mathrm{S}$, the perturbation $\tilde{\chi}$
is spectrally much broader than the soliton $\tilde{\psi}_\mathrm{S}$. Therefore, we can assume that $\chi$ will essentially describe radiation, or, in other words, the part of the total wave function which is completely alien to the soliton.
To proceed, we will compute the ensemble average of $P_{\chi}^{\Delta t}=\int|\chi(x,\Delta t)|^2\mathrm{d}x$, that is, the power of the
radiation produced in the time interval $[0,\Delta t]$:
\begin{equation*}
\begin{split}
 \langle P_{\chi}^{\Delta t} \rangle & = \langle \int \int\limits_0^{\Delta t} \int\limits_0^{\Delta t} \eta(x,t)\psi_\mathrm{S}(x) \eta(x,t^{\prime})\psi^*_\mathrm{S}(x)
\mathrm{d}t^{\prime}\mathrm{d}t\mathrm{d}x\rangle \\
& = C(0) \int |\psi_\mathrm{S}(x)|^2 \mathrm{d}x \Delta t = C(0) P_\mathrm{S} \Delta t,
\end{split}
\end{equation*}
where we used \refeq{eq:corr}. On the other hand, \refeq{eq:local_nls} is conservative which dictates that the fraction of power converted
to radiation $P_{\chi}^{\Delta t}$ is lost to the soliton. It is known that the Schr\"odinger soliton can adapt adiabatically to losses by
moving along the family branch towards lower powers~\cite{Cerda:epd:1998}, i.e., the soliton parameter $\lambda(t)$ and power $P_\mathrm{S}(t)\propto\sqrt{\lambda(t)}$
become slowly decreasing in time. If we assume that the radiation, once produced, does not interact anymore with the soliton and disperses quickly,
\refeq{eq:formalpert} becomes valid for any interval $[t,t+\Delta t]$, and we can conclude that
\begin{equation}
\langle P_\mathrm{S}(t+\Delta t) \rangle = \langle P_\mathrm{S}(t)\rangle \left[1 - C(0) \Delta t\right].
\label{eq:powerconservation}
\end{equation}
Here, $P_\mathrm{S}(t)\propto\sqrt{\lambda(t)}$ denotes the power in the solitonic part of the total wave function $\psi(x,t)$.
With $\Delta t \rightarrow 0$ we therefore get
\begin{equation}
\langle P_\mathrm{S}(t) \rangle = P_\mathrm{S}(t=0) e^{-C(0)t},
\quad \langle \lambda(t) \rangle = \lambda(t=0) e^{-2C(0)t}.
\label{eq:solitonevolution}
\end{equation}
Interestingly, we find that noise induced radiation losses of the soliton are described by the same constant $C(0)$ we
already found in the context of the ``random phase shift'' in \refeq{eq:local_mf}.
However, the two effects are entirely different: \refeq{eq:local_mf} describes $|\langle\psi(x,t)\rangle|^2$,
whereas \refeq{eq:solitonevolution} approximates the soliton power related to $\langle|\psi(x-X(t),t)|^2\rangle$, i.e.,  when both random phase shift and random walk play no role.

\begin{figure}
\includegraphics[width=0.99\columnwidth]{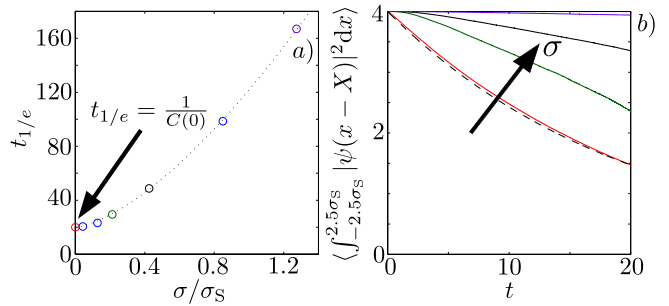}
\caption{(Color online) 
(a) Time $t_{1/e}$ where the soliton power drops below $P_\mathrm{S}(t=0)/e$ versus degree of nonlocality $\sigma/\sigma_\mathrm{S}$.
Numerical simulations confirm the analytical prediction~\refeq{eq:solitonevolution}, $t_{1/e}=1/C(0)$, in the weakly nonlocal limit $\sigma/\sigma_\mathrm{S}\ll1$.
The dotted line represents a guide to the eye $\propto(\sigma/\sigma_\mathrm{S})^{3/2}$.
 The auto-correlation $C(0)=n^2/4\sigma=1/20 $ is kept constant.
(b) The corresponding evolution of the power of the solitons is shown. The correlation lengths used are $\sigma=$ 3, 1, 0.5, and 1/128 for the upper violet, then black, green and finally the lowest red curve, respectively.
In our simulations, ensemble averages are taken over $128$ realizations.
}\label{fig4}
\end{figure}

Figure~\ref{fig4} illustrates the decay of soliton power $P_\mathrm{S}$ due to radiation.
To compute $P_\mathrm{S}$ numerically we integrate over a sufficiently large box around the wave packet,
\begin{equation}
P_\mathrm{S}\approx\int_{-2.5\sigma_\mathrm{S}}^{2.5\sigma_\mathrm{S}}|\psi(x-X)|^2\mathrm{d}x.
\end{equation}
Here we assume that the (delocalized) radiative part $\psi_\mathrm{R}$ of the wave function is negligible in this interval.
In Fig.~\ref{fig4}(a), we show the time $t_{1/e}$ for the ensemble averaged soliton power $P_\mathrm{S}$.
Fig.~\ref{fig4}(b) shows the dynamics of the power decay for different values of the correlation length $\sigma$, and confirms that
our simplified model \refeq{eq:solitonevolution} works well for sufficiently small $\sigma$.
Comparing the times $t_\mathcal{H}$ obtained in the previous section and $t_{1/e}$, we see that $t_\mathcal{H}\ll t_{1/e}$.
Depending on the context and interest, one has to decide which one to use. Clearly, if $t_\mathcal{H}$ is large the soliton will propagate for long times
without significant change. The time $t_{1/e}$ is useful when we are interested in estimating the destruction of the wave packet (i.e., ask the question how long will the soliton survive).

We note that using a similar reasoning one can  estimate radiation losses for an arbitrary initial condition $\psi(x,t=0)$, provided that it is spectrally narrow compared to the noise and
random walk is negligible. Then, it is possible to write down an evolution equation for the so-called mean field $\psi_\mathrm{MF}$, which fulfills
\begin{equation}
\left[ i\partial_t + \partial_{xx} + |\psi_\mathrm{MF}|^2 + i \frac{C(0)}{2} \right] \psi_\mathrm{MF}=0,
\label{eq:local_mean_field}
\end{equation}
and can be connected to the wave function $\psi$ via $|\psi_\mathrm{MF}(x,t)|^2\approx\langle|\psi(x,t)|^2\rangle$. We checked the validity of \refeq{eq:local_mean_field} numerically for various
non-solitonic initial data $\psi(x,t=0)$. Using perturbative variational techniques~\cite{Cerda:epd:1998}, it is simple to show
that the solitons in \refeq{eq:local_mean_field} decay with the same rate as those  found in~\refeq{eq:solitonevolution}.

\subsubsection{White noise ($\sigma\rightarrow0$)}

Before moving over  towards nonlocal solitons, let us discuss the limit of purely white spatial noise, i.e., when $\sigma\rightarrow0$.
In this case the spatial noise term in \refeq{eq:model1}
has to be interpreted appropriately, depending on the underlying physics.
In principle, as $\sigma$ represents a relevant physical length scale~\footnote{This
length scale could be, for example, the mean distance of the molecules in a nematic liquid crystal. },
one has to make sure that it is resolved numerically by choosing a sufficiently fine mesh with spatial step size $\Delta x\ll\sigma$.
However, this may require high computational costs for small $\sigma$. For example, in numerical simulations employing $\sigma=1/128$ we have to use at least $\Delta x \approx 10^{-3}$.
To reduce computational costs, one
could say that this noise is effectively $\delta$ correlated.
Then, according to \refeq{eq:autocorrelation}, in the limit $\sigma\rightarrow0$ and fixed coupling strength $n$ (i.e., keeping the strength of the random walk  constant)
radiation losses $C(0)$ in \refeq{eq:solitonevolution} formally go to infinity. From a physical point of view this is not a problem because $\sigma$
may be small but never actually zero. However, in the case that we do not (or can not) resolve $\sigma$, e.g., in a macroscopic approach, the noise can be considered to be formally $\delta$ correlated, $R(x)=\delta(x)$ and $C(x)=n^2\delta(x)$.

\begin{figure}
\includegraphics[width=0.99\columnwidth]{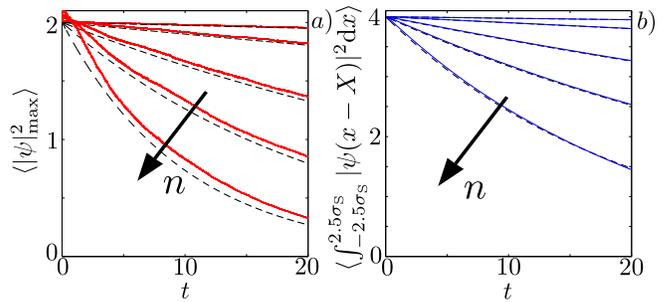}
\caption{(Color online) 
(a)  Evolution of the ensemble averaged peak intensity $\langle|\psi|^2_\mathrm{max}\rangle$
(solid red lines) compared to analytical predictions
\refeq{eq:solitonevolution} (dashed black lines) for  $\delta$ correlated white noise.
The coupling strength is $n=0.005$, 0.01, 0.02, 0.03,  and $\sqrt{0.05\Delta x}$ from the top to the lowest curve, respectively; the spatial step size $\Delta x=80/2048$ is kept constant.
The last value $n=\sqrt{0.05\Delta x}$ is chosen such that it
yields the same decay rate $C(0)$ for the soliton power as
the case $\sigma=1/128$ in~\reffig{fig4}.
(b) The corresponding evolution of the power of the solitons is shown.
In our simulations, ensemble averages are taken over $128$ realizations.
}\label{fig5}
\end{figure}

In our numerical scheme $\delta$-correlation means that $C(0)=n^2/\Delta x$.
Provided that the wave function $\psi$ is sufficiently resolved on the mesh with step size $\Delta x\gg\sigma$, numerical simulations using
$\delta$ correlated noise are supposed to mimic those employing the actual response $R(x)$.
Figure~\ref{fig5} shows simulation results for $\delta$ correlated noise.
Predictions obtained from \refeq{eq:solitonevolution} [or, alternatively, from the mean field \refeq{eq:local_mean_field}] are in excellent agreement with the numerical results. Thus,
as far as radiation losses are concerned,  to mimic an actual response $R(x)$ with finite $\sigma$, we have to choose the same value for the auto-correlation $C(0)$. This essentially means that we have to use an effective coupling strength $n_{\rm eff}\propto \sqrt{\Delta x}$ for the $\delta$ correlated noise. Indeed, comparing the curves for $\sigma=1/128$ in~\reffig{fig4} with those in \reffig{fig5} where $n=\sqrt{0.05\Delta x}$ [i.e., $C(0)=1/20$],
we see that it is actually possible to substitute noise with extremely small correlation length by $\delta$ correlated noise and thus obtain the
same results with much less computational costs. However, by doing so we sacrifice
the accurate description of the random walk of the solitons: for a coupling strength $n_{\rm eff}\propto \sqrt{\Delta x}$ random walk becomes mesh dependent and vanishes for $\Delta x \rightarrow 0$ [see \refeq{eq:randomwalk}].

Obviously, here we have to make a choice of which effect we want to describe correctly and independently of the step size $\Delta x$ in the $\delta$ correlated case.
On one hand, if we choose the random walk to be grid independent ($n^2$ fixed), $C(0)$ and therefore radiation losses become grid dependent
and go to infinity for $\Delta x \rightarrow 0$. This strategy was used by the authors in Ref.~\cite{Hamaide:1991}. On the other hand, if we want
to describe radiation losses correctly, we have to fix the autocorrelation $C(0)$ to the value
of its counterpart for the original correlated noise.

\subsection{Effect of nonlocality of nonlinear potential}\label{sec:nonlocal}

\begin{figure}
\includegraphics[width=0.99\columnwidth]{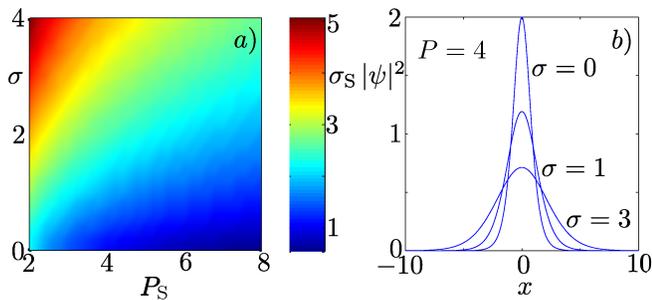}
\caption{(Color online) (a) Dependence of the $1/e$ width $\sigma_\mathrm{S}$ of the nonlocal soliton as a function of power $P_\mathrm{S}$ and degree of nonlocality $\sigma$. 
(b) Exemplary intensity profiles  of nonlocal solitons for different values of $\sigma$ and fixed power $P_\mathrm{S}=4$.
}\label{nonlocal_solitons}
\end{figure}

The stabilizing effect of nonlocal nonlinearities with respect to collapse~\cite{Turitsyn:tmf:85,Bang:pre:2002,Lushnikov:PRA:2010,Maucher:nonlinearity:2011},
perturbations of the initial soliton profile~\cite{Bang:pre:2002}, and even support of higher-order solitons~\cite{Mihalache:PRL:2002,Briedis:opex:2005,Lashkin:pra:2007,Skupin:pre:2006,Maucher:2010:PRA}
has been extensively discussed in the literature.
Based on these results  one might be tempted to expect that the nonlocal character of the
self-induced nonlinear potential itself will be sufficient to weaken the
destabilizing effect of $\delta$ correlated random perturbations on the wave function by spatially
smoothing out the random perturbation of nonlinearity. 
To clarify this issue we will analyze in this ection the following nonlocal model
\begin{equation}
i\partial_t\psi + \partial_{xx} \psi+ \left[  R*|\psi|^2  \right] \psi + \epsilon \psi = 0,
\label{eq:nonlocal_nls_delta_noise}
\end{equation}
that is, opposite to the model considered in the previous subsection, the nonlocal response function $R$ gets convoluted (indicated by $*$) with  the nonlinearity $|\psi|^2 $ and \emph{not} with the white noise $\epsilon(x,t)$.
Thus, the nonlocality of the system will affect  only  the nonlinear self-induced potential without modifying the noise term which will remain $\delta$ correlated. On physical grounds such a model could  describe  nonlinear  beams propagating in an optical waveguide with random perturbation of its parameters (e.g. width or refractive index).
In the case of a nonlocal nonlinearity soliton solutions are no longer available analytically [cf.\ \refeq{eq:soliton_solution} for the local case],
so we have to resort to numerically computed profiles. Figure~\ref{nonlocal_solitons} illustrates the dependence of the width $\sigma_\mathrm{S}$ of nonlocal solitons on degree of nonlocality $\sigma$ and soliton power $P_\mathrm{S}$.

As in the previous section, it turns out that
the evolution  of the average peak intensity of the soliton $\langle|\psi_\mathrm{S}|^2\rangle$ can be accurately  described
by a mean field quantity $|\psi_\mathrm{MF}|^2\approx\langle|\psi|^2\rangle$, and the mean field
amplitude $\psi_\mathrm{MF}$ is governed by the following equation
\begin{equation}
\left[ i\partial_t + \partial_{xx} + R*|\psi_\mathrm{MF}|^2 + i \frac{C(0)}{2} \right] \psi_\mathrm{MF}=0,\label{eq:nonlocal_mean_field}
\end{equation}
in complete analogy to \refeq{eq:local_mean_field}.
This finding  indicates  that  loss of power (or particle number) and subsequent decay of the effect of self-trapping of the soliton solely depends on $C(0)=n^2/\Delta x$.
In particular, for a
given noise level all solitons exhibit {\em the same rate of power loss} independently of the
degree of nonlocality of the nonlinear self-trapping potential, which is somehow counter-intuitive
since in many other situations a stabilizing effect of nonlocal nonlinear potentials has been observed.

Detailed numerical investigations (not shown) confirm that nonlocality
of the nonlinear response as represented in \refeq{eq:nonlocal_nls_delta_noise} has no effect on the
decay of solitons in the presence of white noise. In fact, the only difference between local and nonlocal nonlinear
response is, in this respect, the modification of the soliton transverse profile.

\subsection{Full nonlocal model}\label{sec:full}

\begin{figure}
\includegraphics[width=0.99\columnwidth]{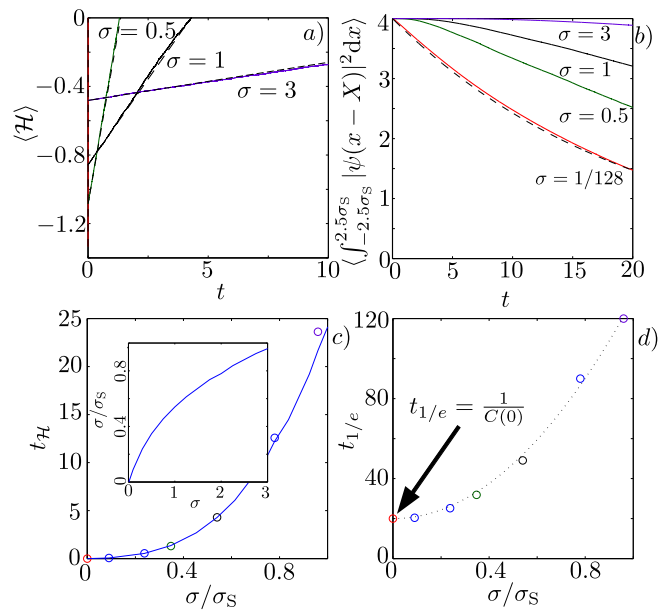}
\caption{(Color online) Soliton dynamics in nonlinear nonlocal model with additive noise \refeq{eq:model1}.
(a)  Numerically found evolution of the averaged Hamiltonian $\langle\mathcal{H}\rangle$ [\refeq{hamnonlocal}]
(solid lines) compared with analytical predictions from
Eq.~(\ref{eq:hamiltonian}) (dashed lines).
The nonlocal length is $\sigma=3$, 1, 0.5, and $1/128$ as indicated. The auto-correlation $C(0)=n^2/4\sigma=1/20 $ is kept constant.
(b)  Corresponding evolution of the power of the solitons.
(c)  Analytically  predicted $t_\mathcal{H}$ (dashed black line) as a function of $\sigma$.
Numerical values are indicated by circles, where the colored circles correspond to the colors used in (a) and (b). The inset in (c) shows the functional dependency of the relative correlation length $\sigma/\sigma_{\mathrm{S}}$ on the nonlocal length $\sigma$ for $P_\mathrm{S}=4$.
(d) Soliton lifetime  $t_{1/e}$ where the soliton power drops below $P_\mathrm{S}(t=0)/e$ versus degree of nonlocality $\sigma/\sigma_\mathrm{S}$.
Numerical simulations confirm the analytical prediction~\refeq{expnonlocal}, $t_{1/e}=1/C(0)$, in the weakly nonlocal limit $\sigma/\sigma_\mathrm{S}\ll1$. The dotted line represents a guide to the eye $\propto(\sigma/\sigma_\mathrm{S})^2$.
In our simulations, ensemble averages are taken over $128$ realizations.
}\label{fig6}
\end{figure}

In the complete model \refeq{eq:model1} nonlocality affects both the self-induced nonlinear potential and the spectrum of the noise. In
the last sections, these two aspects have been considered separately.
We will now discuss the soliton dynamics following from the full nonlocal model of nonlinear media \refeq{eq:model1}.
In the following consideration we fix the initial power of the soliton $P_\mathrm{S}(t=0)=4$. Then, the width of the initial soliton $\sigma_\mathrm{S}$ is just a function of the nonlocal length $\sigma$, which is depicted in the inset of \reffig{fig6}(c).
The time evolution of the analogous Hamiltonian for different values of $\sigma$
\begin{equation}
 \mathcal{H} = \int |\partial_x \psi|^2 -\frac{1}{2}|\psi|^2R*|\psi|^2 \mathrm{d}x, \label{hamnonlocal}
\end{equation}
is shown in \reffig{fig6}(a). It turns out that the analytical results obtained in Sec.~\ref{sec:noise} hold [i.e., with Furutsu-Donsker-Novikov formula we find again \refeq{eq:hamiltonian}].
The only noted difference is that the initial Hamiltonian $\mathcal{H}_0$
(as well as the soliton shape function) is now dependent on $\sigma$, and has to be calculated by using the numerical exact nonlocal soliton profile.
As a consequence, the dashed lines representing the analytical predictions in \reffig{fig6}(a) have different starting points at $t=0$ (solid lines show their numerical verifications).
Moreover, the resulting $t_\mathcal{H}=-4\mathcal{H}_0\sigma^3/n^2P$ is slightly smaller than in the previous case [see \reffig{fig6}(c)].

Finally, let us consider the effect of radiation losses on the nonlocal solitons in Eq.~(\ref{eq:model1}). Figure~\ref{fig6}(b) shows the ensemble averaged soliton power, integrated over an interval of $5\sigma_\mathrm{S}$ centered around the barycenter
$X(t)$, for different values of $\sigma$. In the weakly nonlocal limit, we recover the results found in the previous sections
[compare with \reffig{fig4}(b) and \reffig{fig5}(b)], that is, an exponential decay of the ensemble averaged soliton power
\begin{equation}
\langle P_\mathrm{S}(t) \rangle = P_\mathrm{S}(t=0) e^{-C(0)t}, \label{expnonlocal}
\end{equation}
with decay rate $C(0)$ [cf.~\refeq{eq:solitonevolution}].
Figure \ref{fig6}(d) illustrates the resulting $1/e$ lifetime of the soliton as a function of the degree of nonlocality $\sigma/\sigma_\mathrm{S}$. We also verified that for spectrally narrow arbitrary initial conditions $\psi(x,t=0)$ the mean field equation~(\ref{eq:nonlocal_mean_field}) holds.

\section{nonlocal model with multiplicative noise}\label{sec:nlnoise}
In this section we will briefly discuss the interplay of randomness and nonlocality in a second nonlocal model. We consider a nonlinear
medium described by the following coupled system
\begin{subequations}
\label{eq:model2}
\begin{align}
i\partial_t \psi(x,t) +\partial_{xx}\psi (x,t)+  \rho \psi(x,t) & = 0 \\
-\sigma^2\partial_{xx}\rho+\rho & = (1+\epsilon)|\psi|^2.
\end{align}
\end{subequations}
Unlike the previously
discussed model with additive noise here the noise amplitude is a function of the  soliton amplitude.
In the context of nematic liquid crystals, such a model takes into account fluctuations in the molecular
reorientation due to the optical beam intensity, i.e., in regions with high optical intensity the molecular orientation experiences larger stochastic perturbation than in low intensity regions.
As before,
$\epsilon(x,t)$ is assumed to be a $\delta$ correlated Langevin noise in
both longitudinal and transverse coordinates.
The above coupled system is equivalent to the following nonlocal Schr\"odinger equation
\begin{equation}
 i\partial_t \psi +\partial_{xx}\psi+  R(x) * [(1+ \epsilon(x,t))  | \psi |^2]  \psi = 0,
\label{eq:nls2}
\end{equation}
with the usual exponential kernel $R$.

In the multiplicative model Eq.~(\ref{eq:model2}), the effective strength of the noise depends crucially on the power of the soliton, since the noise term is
multiplied by the intensity.
Similarly to the case of additive noise [Eq.~(\ref{eq:model1})], we can estimate radiative losses in the weakly nonlocal limit. 
Analogously  to \refeq{eq:powerconservation}, one finds that the power of the soliton is governed by
\begin{equation}
 \langle P_\mathrm{S} (t+\Delta t) \rangle = \langle P_\mathrm{S}(t)\rangle - \frac{n^2\Delta t}{4\sigma} \int |\psi_\mathrm{S} (x,t)|^6 dx.
\end{equation}
Using the analytical soliton profile,~\refeq{eq:soliton_solution} (justified only  in case when $\sigma \ll \sigma_\mathrm{S}$)
we find the evolution equation for the soliton parameter
\begin{equation}
 \frac{\mathrm{d}\sqrt{\lambda}}{\mathrm{d}t} = -\frac{8n^2}{15\sigma} \sqrt{\lambda^5},
\end{equation}
which can be solved analytically with the initial condition $\lambda(0)=1$,
\begin{equation}
 \lambda(t) = \frac{15}{\sqrt{480 n^2 t /\sigma + 225}}. \label{lambdamult}
\end{equation}
The corresponding expression for the soliton power can be found as $P=4\sqrt\lambda$.

\begin{figure}
\includegraphics[width=0.99\columnwidth]{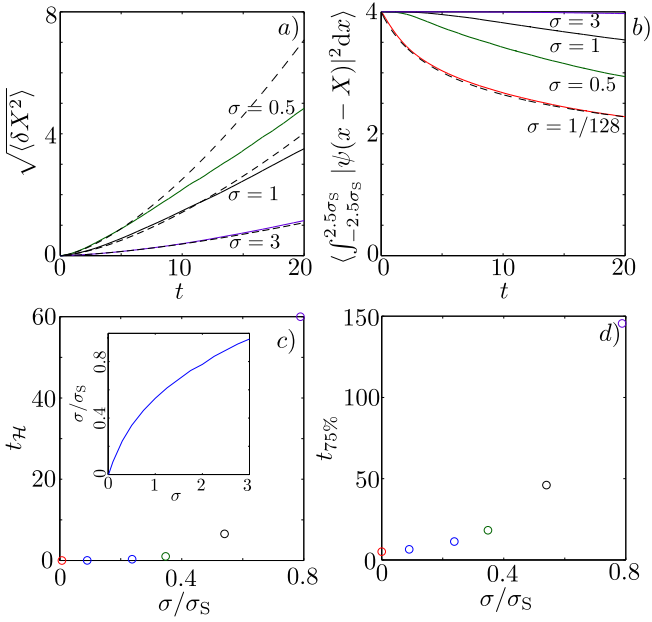}
\caption{(Color online) 
 Nonlinear nonlocal model with multiplicative noise \refeq{eq:model2}.
(a)  Soliton random walk (solid lines) compared to the analytical prediction of \refeq{eq:randomwalk2}  (dashed lines). (b) The corresponding evolution of the power of the solitons is shown.
The correlation length is $\sigma=\infty$, 3, 1, and 1/128 as indicated. The coupling strength $n$ is adjusted such that $n^2/4 \sigma=1/20 $ is kept constant.
(c) Numerical values of $t_\mathcal{H}$ as a function of $\sigma$ are shown.
Colored circles correspond to the colors used in (a) and (b). The inset in (c) shows the functional dependency of the relative correlation length $\sigma/\sigma_{\mathrm{S}}$ on the nonlocal length $\sigma$ for $P_\mathrm{S}=4$.
(d) Numerical values of the time $t_{75\%}$ where the soliton power drops below $0.75P_\mathrm{S}(t=0)$ versus degree of nonlocality $\sigma/\sigma_\mathrm{S}$ are shown.
In our simulations, ensemble averages are taken over $128$ realizations.
}\label{fig7}
\end{figure}

Figure~\ref{fig7}(b) shows the evolution of ensemble-averaged soliton power obtained from numerical simulations. In the case of spectrally broad noise (i.e., small $\sigma$),
we find perfect agreement with our analytical results \refeq{lambdamult}.
To be able to compare with previous findings, $n^2/4\sigma=1/20$ has been fixed at the same value.
The main difference between additive and multiplicative noise is that in the latter case the radiative decay of the solitons is not exponential.
Therefore,
we consider the time $t_{75\%}$, where the ensemble averaged soliton power drops below $75\%$ of $P_\mathrm{S}(t=0)$. As expected, \reffig{fig7}(d) reveals a significant increase of $t_{75\%}$ with increasing degree of nonlocality $\sigma$.

Unlike in the case of additive noise, we cannot write down a mean field equation or give an analytic expression
for the Hamiltonian $\mathcal{H}$ [\refeq{hamnonlocal}], since the effective noise   depends now  on $\psi$ itself in a nontrivial way.
The times $t_\mathcal{H}$ , where $\langle \mathcal{H} \rangle$ becomes zero in our simulations are depicted in \reffig{fig7}(c).
Also the random walk now depends on the evolution of the soliton. Assuming naively, that the soliton profile is conserved upon
propagation,  so that we can take the initial profile as spatial noise filter for all times,
one can find that the random walk of solitons is described  by a formula analogous to Eq.~(\ref{eq:randomwalk})
\begin{subequations}
\label{eq:randomwalk2}
\begin{align}
&\langle \delta X(t)^2 \rangle = \frac{16 n^2 \kappa}{3P_\mathrm{S}^2}t^3 \\
& \kappa  = \iiint \psi_\mathrm{S}(x_1) \left[\partial_{x_1}\psi_\mathrm{S}(x_1)\right]R(x_1-x_3) \label{eq:randomkappa2}\\
& \quad \times \psi_\mathrm{S}(x_2) \left[\partial_{x_2}\psi_\mathrm{S}(x_2)\right] R(x_2-x_3)  |\psi_\mathrm{S}(x_3)|^4\mathrm{d}x_1 \mathrm{d}x_2 \mathrm{d}x_3 \nonumber
\end{align}
\end{subequations}
In case of a large nonlocal length, \refeq{eq:randomwalk2} yields  results which agree with numerics, whereas it fails for smaller  $\sigma$ due to
significant changes in the profile and radiation upon propagation [see \reffig{fig7}(a)].
For better readability of the figure, we do not show the weak random walk of the case $\sigma=1/128$ in \reffig{fig7}(a).

\section{Conclusions}\label{sec:conclusion}

In this paper we investigate systematically the interplay between nonlocality and randomness
for two  different prototypical physical models. Our first model considers the case, where random fluctuations are present independently of the nonlinear wave (additive noise),
whereas in the second model randomness itself is dependent on the wave function (multiplicative noise).
Even though we restrict our analysis to an exponential nonlocal function, our results should hold for arbitrary nonlocal kernels.
We were particularly interested in the stability of solitons.
The stochasticity in the constituent equation leads to radiation which causes the soliton to loose power (or norm).
To characterize the time scales of this radiative power loss, we derived two criteria. The first one involves the Hamiltonian of the noiseless part of the model equations, and gives an estimate for how long the soliton can propagate without changing its shape.
It turns out that the time $t_\mathcal{H}$ of the soliton depends crucially on the degree of nonlocality (i.e., the ratio of the correlation length of the noise over the width of the soliton). Even though we fix the autocorrelation of the noise when changing the degree of nonlocality, which means that the noise amplitude increases for larger nonlocal length $\sigma$, we
find a dramatic enhancement of the soliton lifetime through nonlocality.
Interestingly, however, the main impact on soliton stability comes from the nonlocality-induced increase of the correlation length of the noise and not from the nonlocality of the nonlinearity.
Our second criterion for soliton stability incorporates the fact that a soliton prone to radiative losses can adiabatically transform itself into another member of its family with lower power. As a result, the wave packet remains   confined upon propagation, but gradually decreases its guided power. Therefore, it makes sense to discuss a ``$1/e$-lifetime'' in terms of power. It turns out that the ``$1/e$-lifetime'' also strongly increases with the nonlocal length $\sigma$.
In the case of spectrally very broad noise or very small correlation length compared to the extent of the wave packet, we found analytical expressions quantitatively
describing the loss of power due to radiation in both nonlocal models.
These analytical results are particularly useful to derive upper estimates for decay rates of the solitons.
In particular, the convergence of weakly correlated noise to formally $\delta$ correlated noise has been discussed in depth.
Finally, we addressed the spatial random walk of the solitons modified by radiation as an additional important dynamical feature.

\begin{acknowledgments}
We would like to thank J.~Hope, V.~V.~Konotop and W.~Koch for fruitful discussions.
Numerical simulations were performed at
the National Computing Infrastructure, Australia, Canberra. This project was supported  by the
Australian Endeavour Research Award and the Australian Research Council.
\end{acknowledgments}

\appendix

\section{Derivation of the averaged equation}\label{sec:FN}

For the derivation of the ensemble averaged \refeq{eq:local_mf} we resort to the Furutsu-Donsker-Novikov formula~\cite{Novikov:JETP:1965,Furutsu:1963,Konotop_book,Scott_book}:
\begin{equation}
 \langle\eta \psi\rangle = \int\limits_{-\infty}^t \!\! \int\limits_{-\infty}^{\infty} \langle\frac{\delta \psi(x,t)}{\delta \eta(x^\prime,t^\prime)} \rangle
 \langle\eta(x,t) \eta(x^\prime,t^\prime)\rangle \mathrm{d}x^\prime\mathrm{d}t^\prime
\label{eq:fn}
\end{equation}
Causality in time is reflected by the upper integration limit.
This formula involves the variational derivative of the wave function $\psi$ with respect to the
noise term $\eta$. To compute this quantity we write \refeq{eq:local_nls} in integral form
\begin{equation}
\begin{split}
& \psi(x,t) = \psi_\mathrm{S}(x) \\
& + i\int\limits_{0}^t \left[\partial_{xx} + |\psi(x,t^\prime)|^2 + \eta(x,t^\prime) \right] \psi(x,t^\prime) \mathrm{d}t^\prime,
\end{split}
\end{equation}
where $\psi_\mathrm{S}(x)$ is the initial condition at $t=0$. Then, with $\frac{\delta \eta(x,t)}{\delta \eta(x^\prime,t^\prime)} = \delta(x-x^\prime) \delta(t-t^\prime)$ (see, e.g.,~\cite{Konotop_book}) we find for $0<t^\prime<t$
\begin{equation*}
\begin{split}
& \frac{\delta \psi(x,t)}{\delta \eta(x^\prime,t^\prime)} = i\int\limits_{t^\prime}^t \left[\partial_{xx} +  \eta(x,t^{\prime\prime}) \right]\frac{\delta \psi(x,t^{\prime\prime})}{\delta \eta(x^\prime,t^\prime)}  \mathrm{d}t^{\prime\prime} \\
& + i\int\limits_{t^\prime}^t
\frac{\delta |\psi(x,t^{\prime\prime})|^2 \psi(x,t^{\prime\prime})}{\delta \eta(x^\prime,t^\prime)}\mathrm{d}t^{\prime\prime}+ i\psi(x,t^\prime) \delta(x-x^\prime).
\end{split}
\end{equation*}
Here we used the causality principle again, namely that $\frac{\delta \psi(x,t^{\prime\prime})}{\delta \eta(x^\prime,t^\prime)} = 0$ for $t^{\prime\prime}<t^\prime$.
Finally, in the limit $t^\prime \rightarrow t$ we get
\begin{equation}
\frac{\delta \psi(x,t)}{\delta \eta(x^\prime,t)} = i\psi(x,t) \delta(x-x^\prime). \label{vardef}
\end{equation}
Thus, we find \footnote{Note that the factor $1/2$ in \refeq{halbesdelta} appears due to $\int\limits_0^{\infty}\delta(t)\mathrm{d}t=1/2$}
\begin{equation}
\langle\eta \psi\rangle=i\frac{C(0)}{2}\langle \psi\rangle, \label{halbesdelta}
\end{equation}
and \refeq{eq:local_mf} can be found in straight forward manner by taking the ensemble average of \refeq{eq:local_nls}.

\section{Derivation of the averaged Hamiltonian}\label{sec:mean_ham}

The ascent $\gamma$ of the ensemble averaged Hamiltonian $\langle\mathcal{H}\rangle$ equals the time derivative of the $\langle\mathcal{H}\rangle$. Thus, we have to compute
\begin{align*}
 \partial_t \langle\mathcal{H}\rangle &= \partial_t \langle \int |\partial_x\psi|^2 - \frac{1}{2}|\psi|^4 \mathrm{d}x \rangle \\
 &= \int \langle i\eta \left(\psi^*\partial_{xx}\psi-\psi\partial_{xx}\psi^*\right) \rangle \mathrm{d}x
\end{align*}
Here, $\psi^*$ means the complex conjugate of $\psi$, and for the time-derivatives of the wave function $\psi(x,t)$ we plugged in \refeq{eq:local_nls}. In the next step, we make use of the Furutsu-Donsker-Novikov formula~(\ref{eq:fn}):
\begin{align*}
 \partial_t \langle\mathcal{H}\rangle &= i \int\limits_{-\infty}^{t} \int\limits_{-\infty}^\infty \int\limits_{-\infty}^\infty \left\{ \langle\frac{\delta \left[\psi^*(x,t)\partial_{xx}\psi(x,t)\right]}{\delta \eta(x^\prime,t^\prime)}\rangle  \right. \\
& \!\!\! \left. - \langle\frac{\delta \left[\psi(x,t)\partial_{xx}\psi^*(x,t)\right]}{\delta \eta(x^\prime,t^\prime)}\rangle \right\} \langle \eta(x,t) \eta(x^\prime,t^\prime)\rangle \mathrm{d}x\mathrm{d}x^\prime \mathrm{d}t^\prime \\
& =  \frac{i}{2}\int\limits_{-\infty}^\infty \int\limits_{-\infty}^\infty \langle\frac{\delta [\psi^*\partial_{xx}\psi-\psi\partial_{xx}\psi^*]}{\delta \eta(x^\prime,t)}\rangle C(x-x^\prime) \mathrm{d}x\mathrm{d}x^\prime
\end{align*}
In the second step, we performed the time-integration over $t^\prime$. By evaluating the variational derivatives \refeq{vardef} and integrating by parts we read
\begin{align*}
\partial_t \langle\mathcal{H}\rangle & = \frac{1}{2}C(0)\langle\int \psi^* \partial_{xx} \psi + \textrm{c.c.}~\mathrm{d}x\rangle \nonumber\\
& \!\!\! -\frac{1}{2}\langle\iint \psi^* \delta(x-x^\prime)\partial_{xx} \left[C(x-x^\prime)\psi \right] + \textrm{c.c.}~\mathrm{d}x^\prime\mathrm{d}x\rangle.
\end{align*}
Most of the integrals above turn out to be zero, which can again be seen by integration by parts, and we find
\begin{equation}
 \partial_t\langle\mathcal{H}\rangle = -\frac{d^2 C(x)}{d x^2}\,\bigg|_{x=0}\langle\int|\psi(x,t)|^2\mathrm{d}x\rangle.
\label{eq:mean_ham}
\end{equation}

\section{Numerical methods}\label{sec:numerics}
To solve nonlinear Schr\"odinger-type equations perturbed by randomness, we use a semi-implicit method in the interaction picture described by the authors of Refs.~\cite{Drummond:CPC:1983,Hope:xmds:2008}.
Starting from time $t_0$, one performs a half step $\Delta t/2$ treating
the Laplacian in Fourier space to find an intermediate wave function $\psi_\mathrm{I}(x)$. Next, one solves a fixed point problem in position space at $t_0+\Delta t/2$ by iterating
$$
\psi_\mathrm{F}(x) = \psi_\mathrm{I}(x) + \frac{\Delta t}{2}V\left[x,\psi_\mathrm{F}(x),\epsilon\left(x,t_0+\frac{\Delta t}{2}\right)\right],
$$
where $V$ represents all terms of the nonlinear Schr\"odinger equation except the Laplacian. The resulting ``fixed point'' $\psi_\mathrm{F}(x)$ is then used to perform the full propagation step with regards to $V$ in position space:
$$
\psi_\mathrm{V}(x) = \psi_\mathrm{I}(x) + \Delta t V\left[x,\psi_\mathrm{F}(x),\epsilon\left(x,t_0+\frac{\Delta t}{2}\right)\right].
$$
Finally, we perform the remaining second half-step $\Delta t/2$ in Fourier space treating the Laplacian to obtain $\psi(x,t_0+\Delta t)$.

To avoid reflection of the wave function at the boundaries of the numerical box we implemented absorbing boundary conditions: After each propagation step we multiply $\psi$ by a
filter function (i.e. a function that equals to 1 everywhere apart from the regions close to the boundaries, where it smoothly decays to zero).

The random term is implemented by using Gaussian distributed random numbers generated by a
Box-Muller method~\cite{Box:AnMathStat:1958}. To calculate the average quantities we repeated simulations over  hundreds  of different stochastic realizations. The results presented in the paper have been obtained using 128 realization of the stochastic systems. We checked that this was sufficient to ensure the repeatability and accuracy of calculations.

\bibliography{random}
\end{document}